\documentclass[aps,prb,twocolumn,showpacs]{revtex4} 

\usepackage{longtable}
\usepackage{tabularx}
\usepackage{amsmath}

\usepackage[final]{graphicx}
\usepackage{dcolumn}
\usepackage{bm}
\def\extgra{pdf}

\newcommand{\mm}[1]     {\ifmmode {#1} \else{}${#1}$\fi}
\newcommand{\mmm}[1]    {\ifmmode{}#1 \else{}${#1}$\fi}
\newcommand{\beq}[1]    {\begin{equation} \label{#1}}
\newcommand{ \eeq}{\end{equation}}

\def\fesex{\mm{\rm{ Fe_{2-x}Se_2Cs_{y} } } }
\def\fesex{\mm{\rm{ Cs_{y}Fe_{2-x}Se_2 } } }
\def\fesexx{\mm{\rm{ X_{y}Fe_{2-x}Se_2 } } }
\def\fesetl{\mm{\rm{ Tl_{y}Fe_{2-x}Se_2 } } }

\def\fese{\mm{\rm{ FeSe } } }
\def\usr{\mm{\rm{\mu SR } } }

\def\Tc\mm{T_c}

\begin{document}


\title{\large Iron vacancy superstructure and possible room
temperature antiferromagnetic order in  superconducting $\rm\mathbf{Cs_{y}Fe_{2-x}Se_2}$}


\author{V.~Yu.~Pomjakushin}
\affiliation{Laboratory for Neutron Scattering, Paul
Scherrer
Institut, CH-5232
Villigen PSI, Switzerland}
\author{D.~V.~Sheptyakov}
\affiliation{Laboratory for Neutron Scattering, Paul
Scherrer
Institut, CH-5232
Villigen PSI, Switzerland}
\author{E.~V.~Pomjakushina}
\affiliation{Laboratory for Developments and Methods, PSI, CH-5232
Villigen PSI, Switzerland}
\author{ A.~Krzton-Maziopa}
\affiliation{Laboratory for Developments and Methods, PSI, CH-5232
Villigen PSI, Switzerland}
\author{K.~Conder}
\affiliation{Laboratory for Developments and Methods, PSI, CH-5232
Villigen PSI, Switzerland}
\author{D.~Chernyshov}
\affiliation{Swiss-Norwegian Beam Lines at ESRF, BP220, 38043
Grenoble, France}
\author{V.~Svitlyk}
\affiliation{Swiss-Norwegian Beam Lines at ESRF, BP220, 38043
Grenoble, France}
\author{Z.~Shermadini}
\affiliation{Laboratory for Muon Spin Spectroscopy, Paul
Scherrer Institut, CH-5232 Villigen PSI, Switzerland}


\date{\today}

\begin{abstract}

Neutron and x-ray powder and single crystal synchrotron diffraction of \fesex\ show the presence of superstructure reflections  with propagation vector k=$[{2\over5},{1\over5},1]$ with respect to the average crystal structure {$ I4/mmm$} ($a=4, c=15$\AA). The propagation vector star corresponds to the 5 times bigger unit cell given by transformation {\bf A}=2{\bf a}+{\bf b},  {\bf B}= -{\bf a}+2{\bf b},  {\bf C}= {\bf c}. A solution for the atomic structure is found in the space groups $P4_2/n$ and $I4/m$ with an ordered pattern of iron vacancies corresponding to the iron deficiency $x=0.29$ and Cs stoichiometry $y=0.83$.  The superstructure satellites are more pronounced in the neutron diffraction patterns suggesting that they can have some magnetic contribution. We have sorted out possible symmetry adapted magnetic configurations and found that the presence of AFM ordering with the ordered magnetic moment of Fe with $\simeq 2\mu_B$ does not contradict to the experimental data. However, the solutions space is highly degenerate and we cannot choose a specific solution. Instead we propose possible magnetic configurations with the Fe magnetic moments in $(ab)$-plane or along $c$-axis. The superstructure is destroyed above $T_s\simeq 500$~K by a first-order-like transition.

\end{abstract}


\maketitle

\section{Introduction}
The recent discovery of the Fe-based superconductors has triggered a remarkable renewed interest for possible new routes leading to high-temperature superconductivity. As observed in the cuprates, the iron-based superconductors exhibit interplay between magnetism and superconductivity suggesting the possible occurrence of unconventional superconducting states. Other common properties are the layered structure and the low carrier density. Among the iron-based superconductors \fese\ has the simplest structure with layers in which Fe cations are tetrahedrally coordinated by Se \cite{Hsu2008}. Recently superconductivity at about 30K was found in \fesexx\  for X=K, Cs, Rb \cite{PhysRevB.82.180520,Krzton2010,2010arXiv1012.5525W}. Muon-spin rotation/relaxation ($\mu$SR) experiments evidence that the superconducting state observed in \fesex\ below 28.5(2) K is microscopically coexisting with a magnetic phase with a transition temperature at $T_m=478.5(3)$ K \cite{2011arXiv1101.1873S}. The magnetic phase appears characterized by rather large static iron-moments as the $\mu$SR signal is wiped out below $T_m$. Very recently the AFM order was reported\cite{Bao2011} in superconducting $\rm {K_{0.8}Fe_{1.6}Se_2}$ with $T_{\rm N}=560$~K with the iron magnetic moment 3.31~$\mu_B$.

The average crystal structure of \fesexx\ is the same as in the layered (122-type) iron pnictides with the space group $I4/mmm$\cite{PhysRevLett.101.107006}. Different types of iron vacancy ordering in \fesetl\ were observed long time ago \cite{Sabrowsky,Haggstrom}, including the one with 5 times bigger unit cell. Due to renewed interest to the superconducting chalcogenides many new experimental studies on the vacancy ordering in \fesexx\ (X=K,Tl) have appeared very recently \cite{Fang2010,Wang2011,Zavalij2011,Wang2011,Basca2011,Bao2011}.

In the present paper we report on the observation of superstructure in superconducting ($T_c=28.5$~K) \fesex\ below $T_s\simeq 500$~K and analyze the diffraction data assuming iron vacancy ordering and possible antiferromagnetic ordering of Fe at room temperature. The single crystals used in the present study are the same as in the Refs.~\cite{Krzton2010,2011arXiv1101.1873S}

\section{Samples. Experimental}
\label{exp}

Single crystals of cesium intercalated iron selenides of nominal compositions $\rm {Cs_{0.8}(FeSe_{0.98}})_2$  were grown from the melt using the Bridgman method as described in Ref.~\cite{Krzton2010}. Powder x-ray diffraction was performed using a D8 Advance Bruker AXS diffractometer with CuK$_\alpha$ radiation. For these measurements a fraction of the crystal was cleaved, powdered, and loaded into the low background airtight specimen holder in a He-glove box to protect the powder from oxidation. Differential scanning calorimetry (DSC) experiments were performed with a Netzsch DSC 204F1 system. Measurements were performed on heating and cooling with a rate of 10 K/min using 20 mg samples encapsulated in standard Al crucibles. An argon stream was used during the whole experiment as protecting gas. Neutron powder diffraction experiments were carried out at the SINQ spallation source of Paul Scherrer Institute (Switzerland) using the high-resolution diffractometer for thermal neutrons HRPT \cite{hrpt} ($\lambda=1.866, 1.494$~\AA, high intensity mode $\Delta d/d\geq1.8\cdot10^{-3}$). Refinement of crystal and magnetic structures of powder neutron diffraction data were carried out with {\tt FULLPROF}~\cite{Fullprof} program, with the use of its internal tables for scattering lengths and magnetic form factors. Single crystal diffraction data were collected  at the SNBL beamline BM1A at the ESRF synchrotron in Grenoble (France) with a MAR345 image-plate area detector using $\lambda= 0.6977(1)$~\AA. Intensities were indexed and integrated with {\tt CrysAlis}\cite{crysalis}, empirical absorption correction was made with SADABS \cite{sadabs}, structure refinement with {\tt SHELXL97}\cite{shel}.

\section{Results and discussion}
\label{res}

\begin{figure}[h]
  \begin{center}
    \includegraphics[width=9cm]{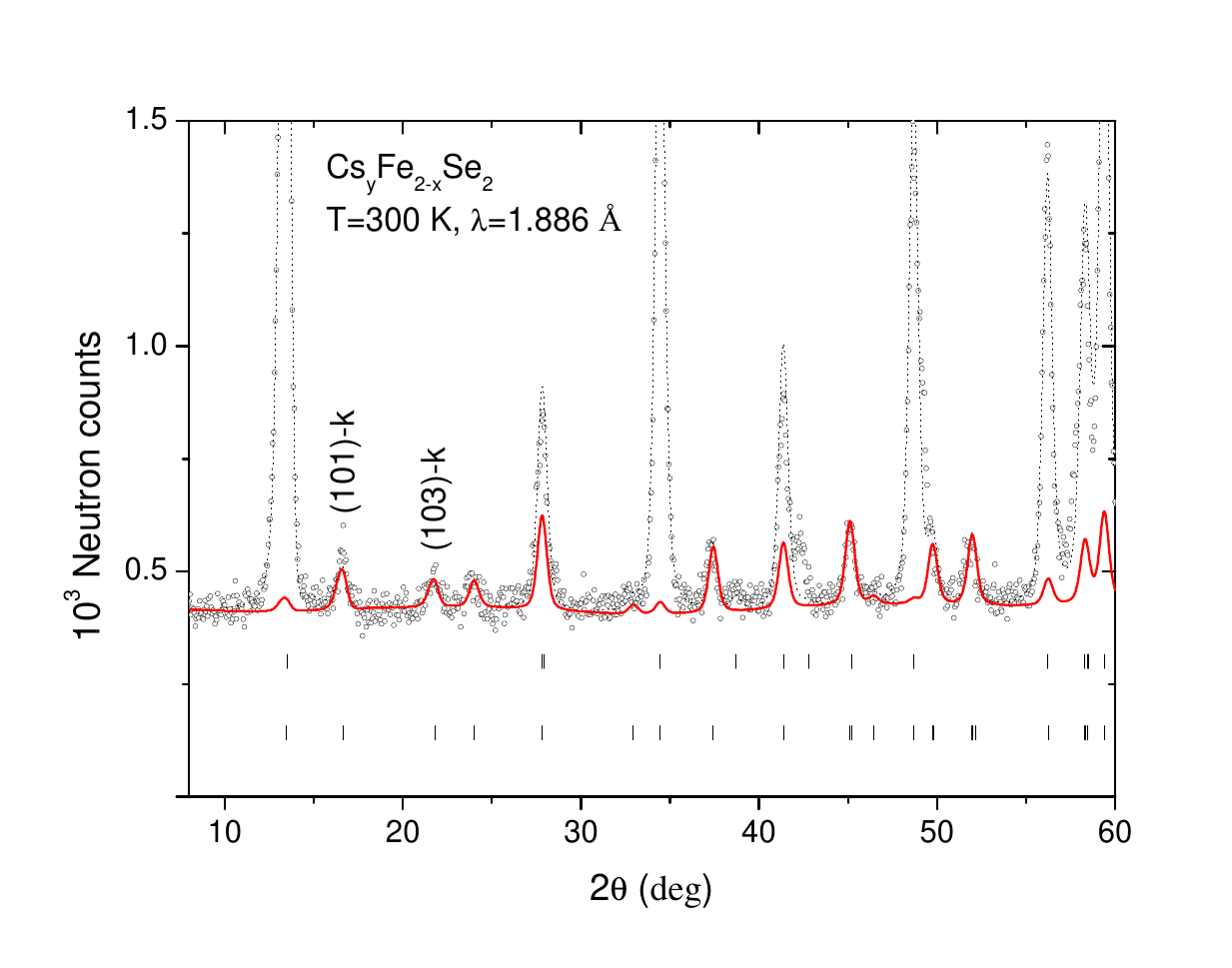} 
  \end{center}

\caption{Fragment of the neutron diffraction pattern of \fesex. Contribution of the superstructure peaks refined in the powder profile matching mode for k=$[{2\over5},{1\over5},1]$ with respect to the average crystal structure ({$ I4/mmm$} with $a=3.96, c=15.29$~\AA\ and the structure parameters fixed by the values from Table~\ref{T1}) is shown by red solid curve. }
  \label{rietv}
\end{figure}

\begin{table}
\caption{Crystal structure parameters refined in the average structure model $I4/mmm$ (no. 139), Fe in $(0,{1\over2},{1\over4})$ (4d), Se in $(0,{0},{z})$ (4e) and Cs in $(0,{0},{0})$ (2a) positions. In the space group I4/m (no. 87) which unit cell is generated by the transformation given in the text the atoms are split in the following way: Cs in $(0,0,0)$ (2a) and $(0.4,{0.8},0)$ (8h), Se in $(0.4,{0.8},-z_{Se})$ (16i) and $({1\over2},{1\over2},-z_{Se}+{1\over2})$ (4e), Fe in $({0.3},{0.6},0.25)$ (16i) and $({1\over2},0,{1\over4})$ (4d), where $z_{\rm Se}$ is z-component in $I4/mmm$. The stoichiometries o-Cs and o-Se are calculated to be in units of the formula \fesex. The data for laboratory x-ray and neutron powder diffraction NPD are given at room temperature, for the synchrotron single crystal (s.c.) experiment at T=536~K in true $I4/mmm$ symmetry above the order-disorder transition.}
 \label{T1}

\begin{center}
\begin{tabular}{lllll}

       &    x-ray   &    NPD        &    NPD,$I4/m$  & s.c. x-ray  \\ 
       &   300~K    &   300~K       &   300~K        & 536~K \\ \hline
 a     & 3.9608(2)  &  3.9614(2)    &  8.8582(3)     & 4.0177(5) \\    
 c     & 15.285(1)  & 15.2873(9)    & 15.2873(9)     & 15.333(4) \\
 z-Se  & 0.3456(4)  &    0.3436(3)  &  -0.343(3)     & 0.3443(3) \\
 o-Cs  & 0.636(13)  & 0.622(24)     & 0.73(1)        & 0.754(9)\\ 
 o-Fe1  &1.48(3)     & 1.49(3)      & 1.50(2)        & 1.66(8)\\
 o-Fe2  &           &               & 0.024(8)        &       \\
B-Fe   &  1.4(3)    &  1.7(1)       &  2.1(1)        & 1.3(2)\\
B-Se   & 4.3(2)     &  3.6(1)       &  3.2(1)        & 1.2(2)\\
B-Cs   & 2.6(3)     &  4.8(5)       &  5.9(5)        & 3.2(2)\\ 
$R_{wp}$,\%& 5.95      & 7.47          & 7.12           & 21 \\
$\chi^2$& 3.6       & 3.3           & 2.86            &  1.4  \\ \hline
\end{tabular}
\end{center}
\end{table}

\begin{figure}[h]
  \begin{center}
    \includegraphics[width=5cm]{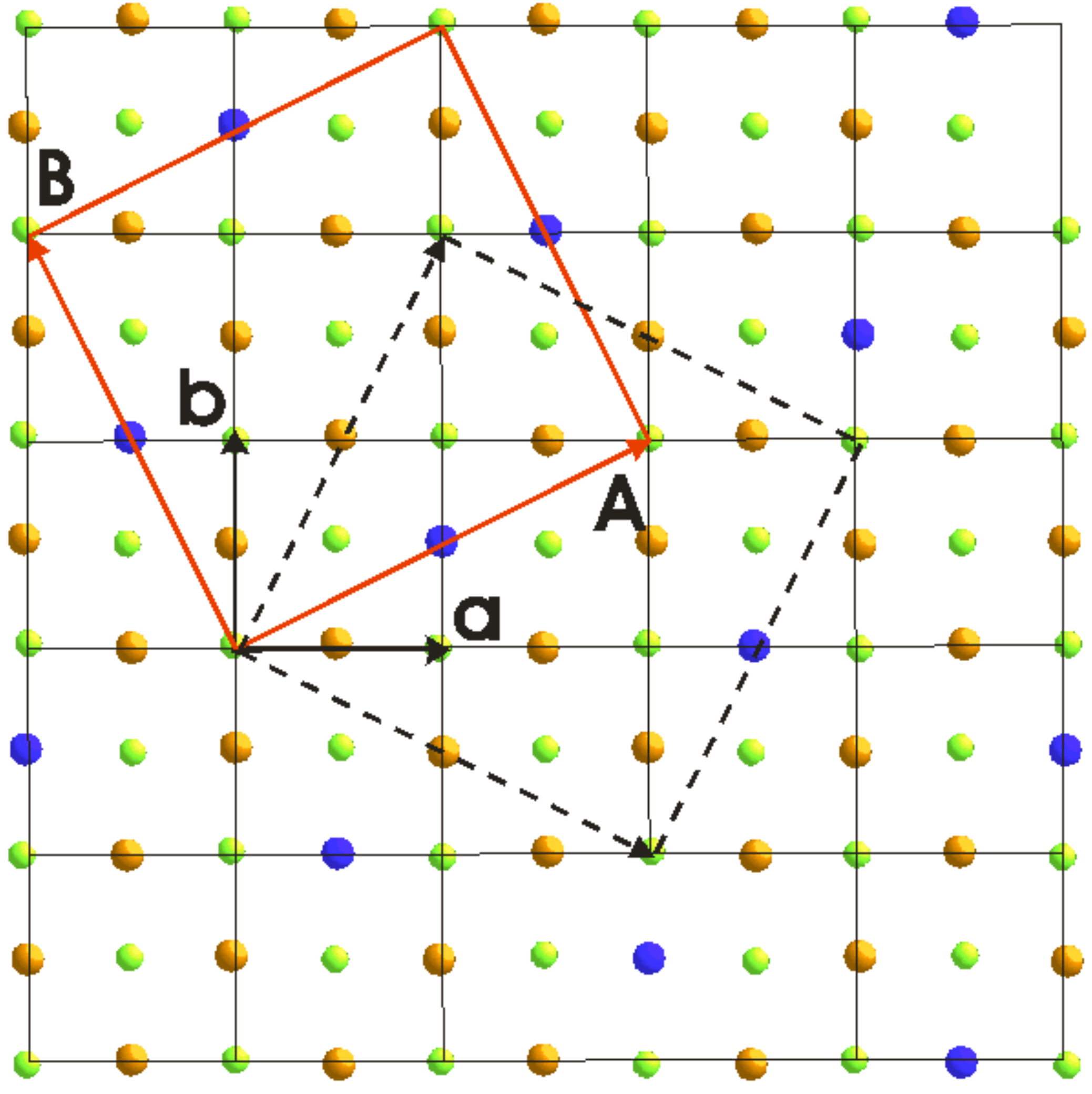}

  \end{center}

\caption{The lattice unit cell transformation to a new bigger tetragonal supercell shown by capital letters. The supercell shown by solid red lines corresponds to the propagation vector star generated by {\bf k}$_1=[{2\over5},{1\over5},1]$.  The dashed cell shows the twin domain that corresponds to the star generated by {\bf k}$_2=[{1\over5},{2\over5},1]$ The k-vector stars are shown in Fig. \ref{recip}. Iron vacancy ordering pattern in $ab$-plane is shown by blue circles. The brown and green circles show fully occupied Fe and Se positions projected to the $ab$-plane. }
  \label{unitcell}
\end{figure}

\begin{figure}[h]
  \begin{center}
    \includegraphics[width=8cm]{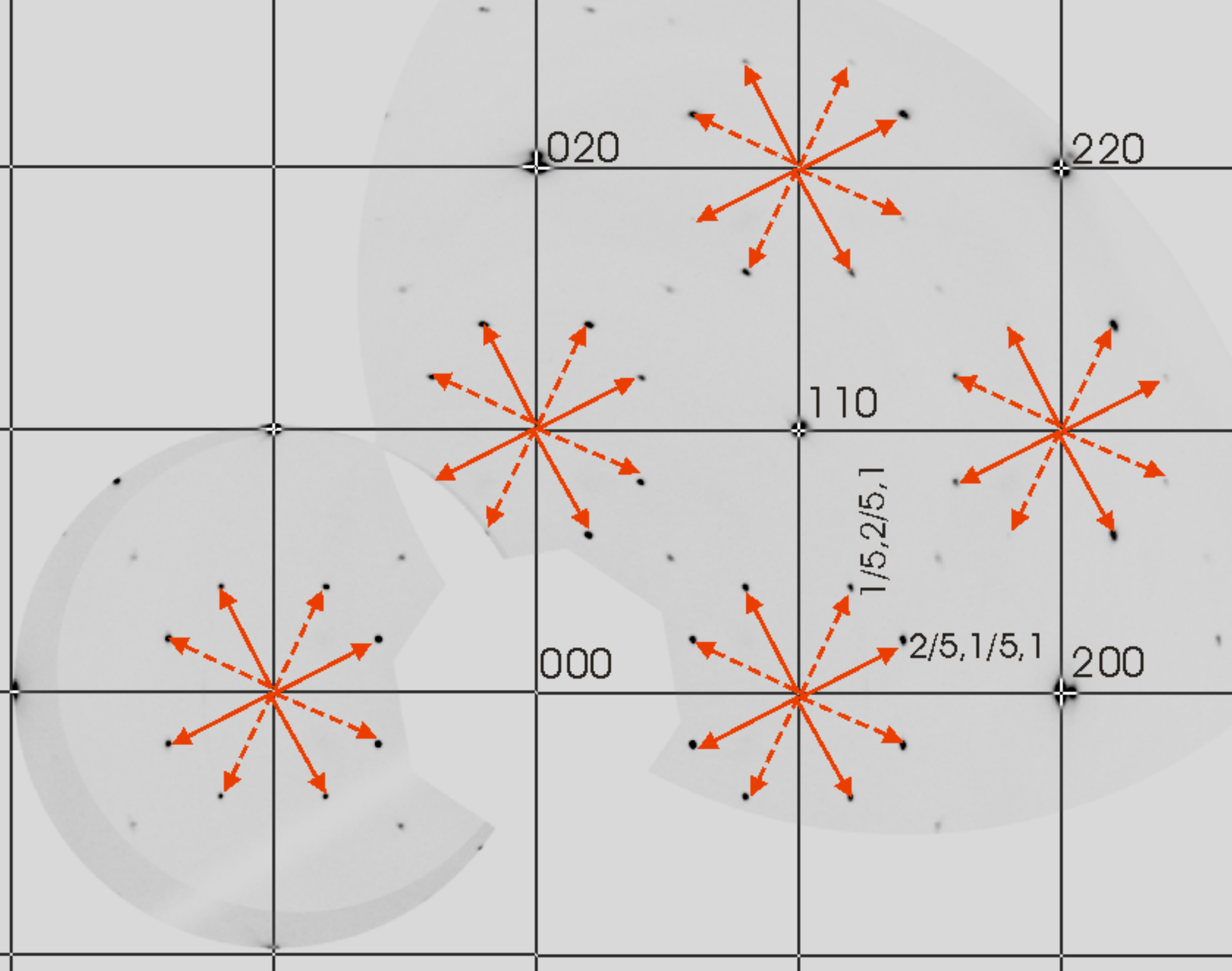}

  \end{center}

\caption{A slice of the reciprocal space showing [hk0] plane. The indexing is given in the average cell ($I 4/mmm$). The satellite reflections are indicated by red arrows for {\bf k}$_1=[{2\over5},{1\over5},1]$ and  {\bf k}$_2=[{1\over5},{2\over5},1]$ by solid and dashed lines, respectively. These two k-vectors correspond to two twin domains shown in Fig.~\ref{unitcell}. }
  \label{recip}
\end{figure}

The average crystal structure can be refined in the standard structure model \cite{Krzton2010}. The iron site occupancy is refined to smaller than unity for both x-ray and neutron diffraction data indicating the presence of the vacancies on the iron sites. The structure parameters refined in this model are presented in Table \ref{T1} for both laboratory x-ray and NPD data. The neutron diffraction pattern has a set of extra  diffraction peaks that can be indexed with the propagation k-vector k=$[{2\over5},{1\over5},1]$, as shown in Fig.~\ref{rietv}. However the x-ray powder diffraction pattern contains only one clearly visible satellite $({3\over5},-{1\over5},0)$ at $q=1$~\AA$^{-1}$, that allows one to suggest that the satellites seen by neutron might have magnetic contribution. The propagation vector star ${k}=\{[{2\over5},{1\over5},\pm1],[-{2\over5},-{1\over5},\pm1],[{1\over5}, -{2\over5},\pm1],[-{1\over5},{2\over5},\pm1]\}$ corresponds to the new unit cell given by the transformation {\bf A}=2{\bf a}+{\bf b},  {\bf B}= -{\bf a}+2{\bf b},  {\bf C}= {\bf c} (the illustration of the lattice cell transformation is shown in Fig.~\ref{unitcell}). A good refinement of NPD pattern explaining the satellite peaks can be done with the supercell indicated above in the space groups $P 4_2/n$ and $I 4/m$. Using the fixed new atomic positions generated from the average crystal structure ($I4/mmm$) by applying  the above basis transformation and releasing only the site occupancies and $z$-Se, similar as for the average structure, one immediately gets a reasonably good description of the superstructure peaks. Table \ref{T1} shows the atomic positions and details of the refinements. The Fe site splits in two sites in $I4/m$ (no. 87) and in three sites in $P4_2/n$ [no. 86, note to work in the second setting with origin at $-1$ post-matrix translation $({1\over4},{1\over4},{1\over4})$ should be applied]. Both groups give similar quality of the refinements of the single crystal data as we explain below, so we present the results only for $I 4/m$ space group, which is more symmetric with respect to the iron sites. The atomic positions for $I4/m$ space group generated by the above transformation from average space group $I 4/mmm$ are listed in Table \ref{T1}. All ``symmetric'' positional parameters which were generated from the special coordinates of $I 4/mmm$ were fixed in the refinement. The fully occupied Fe1 (16i) site contributes 1.6 of iron stoichiometry in \fesex. If both site occupancies are refined, Fe1 gets almost maximal value, whereas Fe2 (4d) site occupancy is close to zero, as shown in the Table \ref{T1}.  The iron vacancy pattern looks like as shown in Fig.~\ref{unitcell}.

\begin{table}
\caption{
Crystal structure parameters refined in the space group
$I4/m$ (no. 87) from the synchrotron single crystal experiment at room temperature. 
The Wyckoff site symmetry positions are the same as described in the caption of Table \ref{T1}.  
The refined stoichiometry is ${\rm Cs_{0.83(1)}Fe_{1.71(1)}Se_2}$, The Fe1 site occupancy was fixed to 1. 
Anisotropic atomic displacement parameters $U_{ij}$ are in $\rm \AA$ multiplied by $10^3$.
 $U_{eq}$ is defined as one third of the trace of the orthogonalized $U_{ij}$ tensor.
Totally 40 parameters were refined using 6286 reflections 539 of which are independent. Final R factors are R1 = 0.0860, wR2 = 0.1955 [$I > 2\sigma(I)$] and R1 = 0.0940, wR2 = 0.2105 (all data).
}
 \label{T2}

\begin{center}
\begin{tabular}{llll}

    & x y z                  & $U_{iso}$      & occ   \\ \hline
Cs1 & 0.0000  0.0000  0.0000       & 78(2)   & 0.911(14) \\
Cs2 & 0.4041(2) 0.8057(2) 0.000 & 75(2) & 0.81(1) \\
Se1 & 0.3924(2)  0.7987(2)  0.6551(2)  & 55(2) & 1 \\
Se2 & 0.5000 0.5000 0.1488(2)     & 54(2) & 1 \\
Fe1 & 0.3014(1) 0.5938(1) 0.25165(6) & 59(2) & 1 \\
Fe2 & 0.5000 0.0000 0.2500                 & 56(9) & 0.27(2) \\ 
\end{tabular}

\begin{tabular}{lllllll}
 &$U_{11}$  & $U_{22}$  & $U_{33}$  &$U_{23}$    & $U_{13}$     &$U_{12}$ \\ \hline
Cs1 &92(2)  & 92(2)  & 49(4)  &0    & 0      &0 \\
Cs2 &84(2)  & 93(2)  & 49(3)  &0    & 0      &3.9(7)\\ 
Se1 &61(2)&   59(2)  & 45(2)  &2.1(5)& 1.1(5)  &1.6(5)\\ 
Se2 &58(2)&   58(2)&   44(3)  &0    & 0      &0\\ 
Fe1 &63(3)  & 62(2)  & 51(4)  &0.8(7)& -1.1(6) & -0.1(7)\\ 
Fe2 &62(10) & 62(10) & 45(16) &0    & 0      & 0 \\

\end{tabular}
\end{center}
\end{table}

\begin{figure}[h]
  \begin{center}
    \includegraphics[width=9cm]{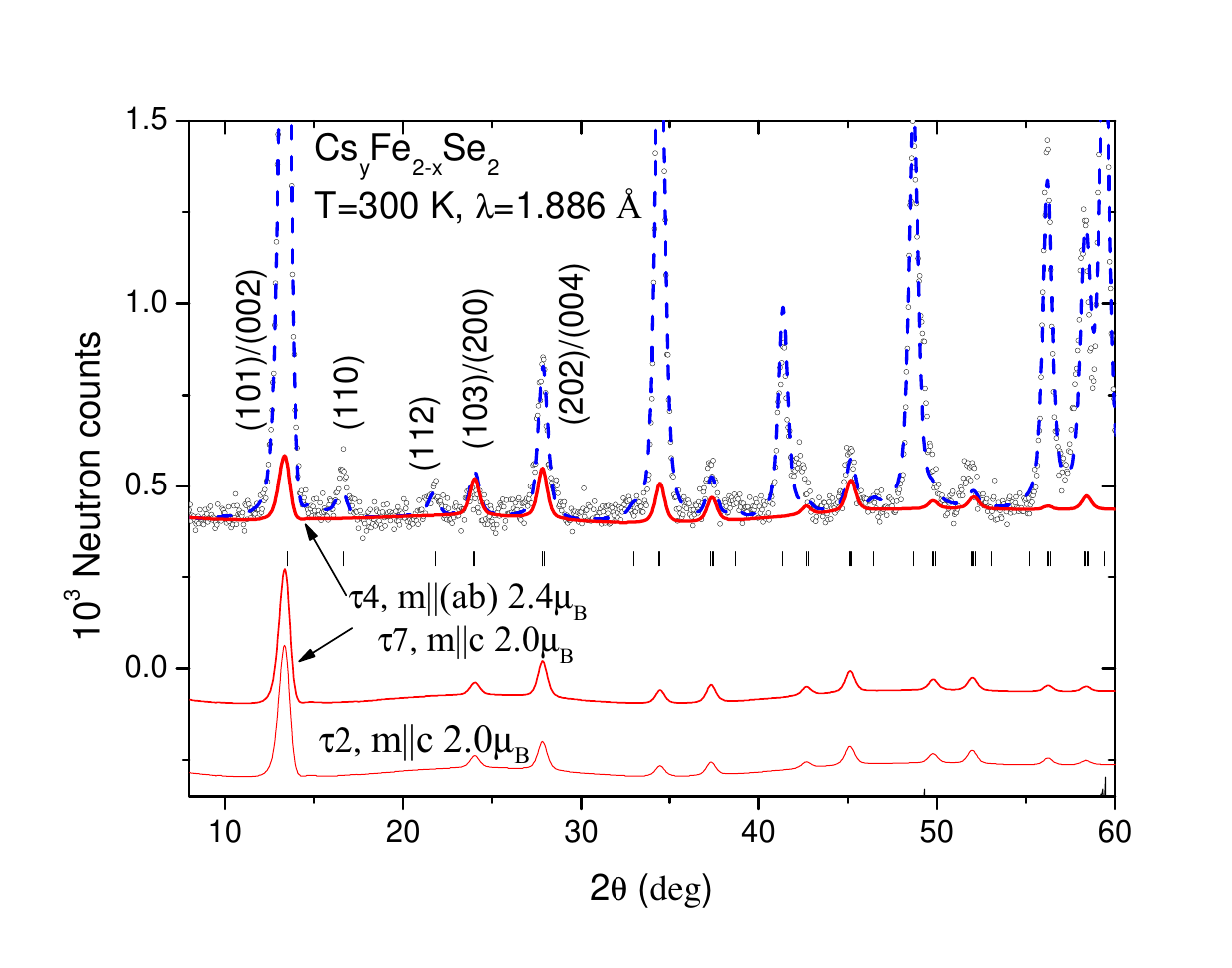} 

  \end{center}

\caption{Fragment of the experimental NPD pattern and the refined profiles for three magnetic models given by symmetry adapted basis functions in $I4/m$ space group. The full profile is shown by dashed blue line, the solid red lines show magnetic contributions to the profile. The curves for $\tau_7$ and $\tau_2$ are shifted along y-axis for better visibility. See text for details. }
  \label{magdif}
\end{figure}

\begin{figure}[h]
  \begin{center}
    \includegraphics[width=9cm]{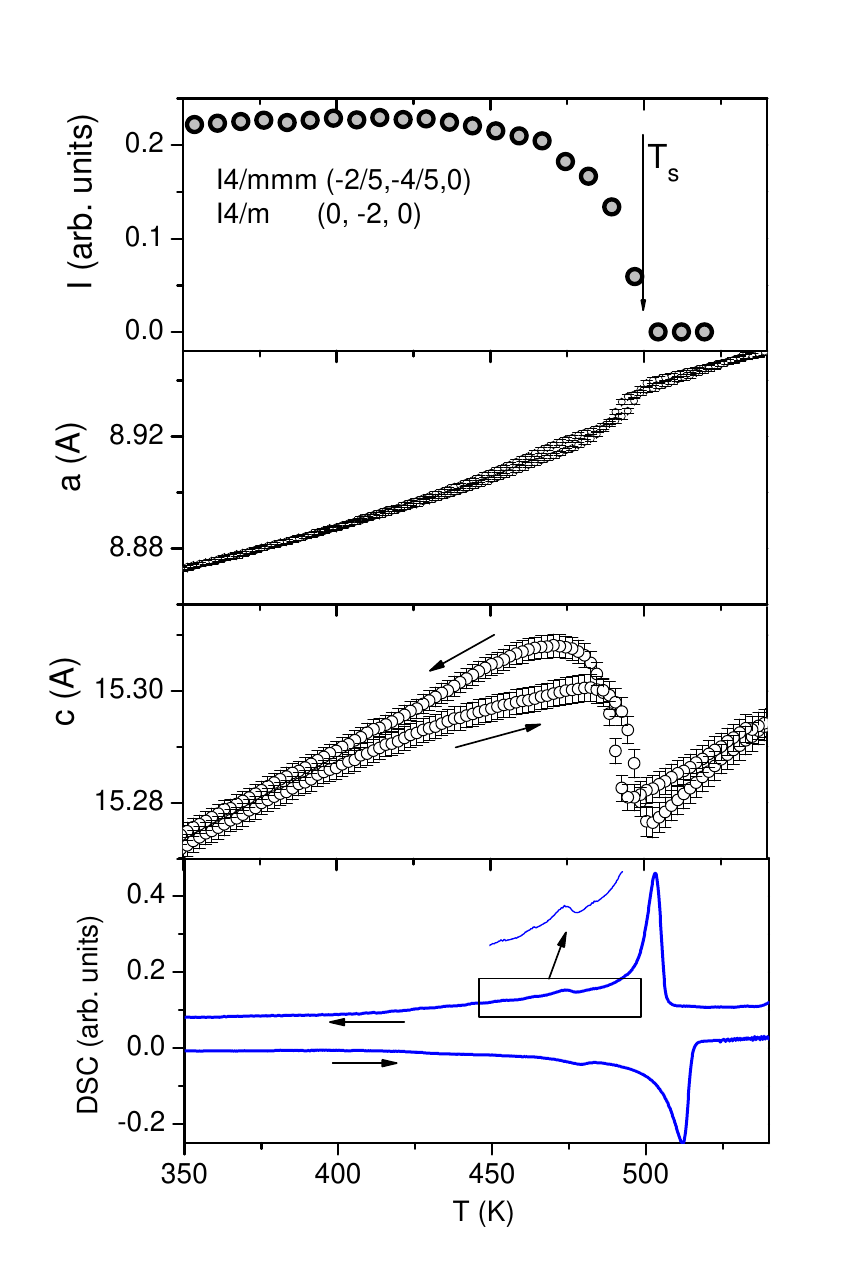} 
  \end{center}

 \caption{
Integrated intensity of (0,-2,0) superstructure satellite, lattice constants obtained from the synchrotron measurements and differential scanning calorimetry (DSC) signal as a function of temperature. The superstructure satellite intensity was obtained from the single crystal measurement on heating. The lattice constants were refined from powder diffraction synchrotron data both on heating and cooling. 
}
  \label{tdep}
\end{figure}

\begin{table}

\caption{Irreps $\tau_i, i=1..8$ of the space group $I4/m$ (no. 87)
for k-vector $k=0$. Numeration of the irreps is in accordance with
Kovalev book ($k=k_{14}$, table $T121$) \cite{Kovalev}. $\tau_4=\tau_3\tau_2$,
$\tau_6=\tau_5\tau_2$, $\tau_8=\tau_7\tau_2$.  The basis functions
$\psi$ for the spin of iron atom in general (16i)-position for
$\tau_4$ (two in  $(ab)$-plane) and for $\tau_7$ (one along $c$-axis)
are given for illustration of possible magnetic structures.}
 \label{T3}

\begin{center}
\begin{tabular}{lrrrrrrrr}

$\tau$,$\psi$& $h_1$     & $h_{14}$ & $h_4$ & $h_{15}$ & $h_{25}$
&$h_{38}$ &$h_{28}$ &$h_{39}$  \\
             &   $1$     & $4^+_z$ & $2_z$ & $4^-_z$ & $-1$ &$-4^+_z$
&$m_z$ &$-4^-_z$  \\\hline
$\tau_2$ &  1&             1 &          1 &           1  &         
-1&                -1&         -1&           -1\\

$\tau_3$ &  1&              i&          -1&           -i&            
1&                 i&         -1&           -i\\

$\tau_5$ &  1&             -1&           1&           -1&            
1&                -1&          1&           -1\\

$\tau_7$ &  1&             -i&          -1&            i&            
1&                -i&         -1&            i\\ \hline
$\psi_4,ab$&  1,0&  0,-i&  1,0&   0,-i&             -1,0&                -1,0&         0,i&            0,i\\
$\psi_4,ab$&  0,1&   i,0&  0,1&   i,0&             0,-1&              0,-1&         -i,0&            -i,0\\
$\psi_7,c$&  1&  i&               -1&   -i&             1&            
   i&         -1&            -i\\
\end{tabular}
\end{center}
\end{table}

Since both vacancy superstructure and possible magnetic structure with $k=0$ contribute to the same neutron Bragg peaks one needs the reliable crystal structure data to disentangle possible magnetic contribution. For this purpose
 several data sets were collected in the single crystal x-ray synchrotron experiment at room temperature and at 536K above $T_m$ and $T_s$. In addition, a limited slice of reciprocal space around $(-{2\over5},-{4\over5},0)$ satellite was collected at heating to identify the transition. Figure \ref{recip} shows a slice of the reciprocal space near $[hk0]$ plane. The superstructure reflections belonging to two domains as shown in the figure can be easily identified. The angle between the domains amounted to 53.2$\rm ^o$ in accordance with the drawing of Fig.~\ref{unitcell}. Note that the extra peaks are centered around forbidden nodes, because they are satellites of the Bragg peaks from the adjacent $[hk1]$ and $[hk\!-\!1]$ planes. The refined structure parameters together with the reliability factors are given in the Table \ref{T2}.  Due to the strong absorption correction effects the atomic displacement parameters ADP can have an overall systematic shift. The refinement of the single crystal data in $P4_2/n$ space group gives slightly worse, but still acceptable reliability factors that we list here for completeness R1 = 0.0860, wR2 = 0.2131 [$I > 2\sigma(I)$] and R1 = 0.1283, wR2 = 0.2885  (all data). One can notice additional diffraction spots at $({n\over2},{m\over2},0)$ in Fig.~\ref{recip}. These spots are actually a projection of satellite rods at $({n\over2},{m\over2},l)$. The in-plane propagation vector is equal to $[{1\over2},{1\over2}]$ in both average cell and supercell.  We do not have any model to account for this additional superstructure, but it must correspond to a 2D-ordering within the $(ab)$-plane, e.g. an ordering of vacancies in Cs layers without correlations between the layers along $c$-axis.  

Using the structure data for $I 4/m$ from the single crystal x-ray experiment we made an attempt to evaluate the magnetic contribution in the NPD data. We assume that only fully occupied Fe1 site has a magnetic moment. All the structure parameters from Table \ref{T2} where fixed in the subsequent NPD refinements. Only overall ADP was introduced to account for the absorption effects in the x-ray experiment. The space group $I 4/m$ has eight one dimensional irreducible representations (irreps) for $k=0$, and all eight irreps enter three times in the magnetic representation for the iron in general (16i) position. The irreps in Kovalev notation \cite{Kovalev} are listed in Table \ref{T3}. There are four complex irreps with Herring coefficient 0 and four real irreps that correspond to the respective Shubnikov groups of $I 4/m$. We sorted out all the irreps and found that there are different magnetic configurations with the moment size about 2$\mu_B$ per iron site that do not contradict to the NPD data. The magnetic R-factors amounted to 17.5-24.5\% for different irreps. For the illustration of the magnetic contribution we show in Table \ref{T3} and in Fig.~\ref{magdif} two ``orthogonal'' magnetic models. For $\tau_4$ we choose the basis functions with the moments in the (ab) plane, whereas for $\tau_7$ the moments are parallel to $c$-axis. Both models have practically the same magnetic Bragg R-factors 18.5 and 17.5\%, respectively. We note that the magnetic moment sizes on the Fe1  sites are not restricted to be the same by symmetry for complex irreps even if we consider the basis function only along one axis. For instance $\psi_7$ can be multiplied by an arbitrary phase factor $\exp(i\varphi)$ that would result in two different moment values. By choosing the phase $\varphi=\pi/4$ all the moments are constrained to be the same.

The model proposed in Ref.~ \cite{Bao2011} corresponds to $\tau_2$ with the Shubnikov symbol $I4/m'$. Unfortunately, we do not observe an explicit magnetic contributions in (101) Bragg peak as observed in Ref.~ \cite{Bao2011} in K-intercalated \fese. This might be partially due to the fact that for the lattice constants of \fesex\ the (101) and (002) appear at the same scattering angle.  In addition, in our case of \fesex\ the magnetic contribution is not so large. For comparison we show also the contribution of this model ($\tau_2, || c$) to the diffraction pattern (Fig.~\ref{magdif}). One can see that the contributions of both $\tau_7$ and $\tau_2$  models are very similar (there are small differences hardly visible on the figure scale), but the magnetic configurations are different, namely for $\tau_7$ the constant moment configuration corresponding to the operators listed in the table are 1,-1,-1,1,1,-1,-1,1, whereas for $\tau_2$ 1,1,1,1,-1,-1,-1,-1. Probably the $\tau_7$ model would also fit the data of Ref.~\cite{Bao2011}. We would like to stress that the possible solutions are highly degenerate by the values of the R-factors and we cannot choose a specific model on the basis of our experimental data.

Figure \ref{tdep} shows the integrated intensity of the superstructure satellite and the lattice constants as a function of temperature obtained in the single crystal and powder synchrotron diffraction experiments. The intensity gradually disappears with transition temperature $T_s\simeq500$~K, whereas the lattice constants exhibit a pronounced hysteresis indicating a first order phase transition. The unit cell volume is linear and does not have a visible peculiarity in the temperature region shown in Fig.~\ref{tdep}. Interestingly, the $c$-lattice constant shows a decrease by 0.1\% at the transition to the disordered phase. The crystal structure above $T_s$ is well refined in the $I4/mmm$ model (Table \ref{T1}). The DSC signal has two peaks, one large at higher temperature and the second small one at lower temperature, which had been associated with the onset of the magnetic order from \usr-experiment~\cite{2011arXiv1101.1873S}. The large DSC peak seems to be originated from the vacancy order-disorder transition at $T_s\simeq500$~K.

\section*{A{\lowercase{cknowledgements}}}

The authors acknowledge the allocation of the beam time at Swiss-Norwegian beam line (BM1A) of the European Synchrotron Radiation Facility (ESRF, Grenoble, France). Fruitful discussions with A.~Bosak are gratefully acknowledged. The authors thank the NCCR MaNEP project and Sciex-NMS$\rm ^{ch}$ (Project Code 10.048) for the support of this study. The work was partially performed at the neutron spallation source SINQ.

\bibliography{../../refs/refs_general,../../refs/refs_fesex,../../refs/refs_manganites,../../publication_list/publist2007,../../publication_list/publist2006,../../publication_list/publist2005,../../publication_list/publist}

\end{document}